\documentclass{article}
\usepackage{comment}
\usepackage{graphicx}
\usepackage{multirow}
\usepackage{authblk}
\providecommand{\keywords}[1]{\textbf{\textbf{keywords:}} #1}

\author[1,2]{Gizem Gezici}
\affil[1]{Sabanci University, Department of Computer Science and Engineering, Istanbul, Turkey}
\affil[2]{Huawei Turkey R\&D Center, Istanbul, Turkey}
\date{}  
\title{Customising Ranking Models for Enterprise Search on Bilingual Click-Through Dataset}

\begin{document}           
%

%
%

\maketitle              

\begin{abstract}
In this work, we provide the details about the process of establishing an end-to-end system for enterprise search on bilingual click-through dataset. The first part of the paper will be about the high-level workflow of the system. Then, in the second part we will elaborately mention about the ranking models to improve the search results in the vertical search of the technical documents in enterprise domain. Throughout the paper, we will mention the way which we combine the methods in IR literature. Finally, in the last part of the paper we will report our results using different ranking algorithms with $NDCG@k$ where k is the cut-off value.
\end{abstract}

\section{Introduction}
Search engines are ubiquitous. Currently, on average 3.5 billion Google searches are done per day~\cite{InternetLiveStats}. Nonetheless, they are also very crucial in enterprise domain to access documents from various information sources. ``The application of information retrieval technology to information finding within organisations has become known as \emph{enterprise search}~\cite{EnterpriseSearch}". It is reported that employees spend 2.5 hours on average per day searching for information~\cite{SpendTime}. According to Transparency Market Research~\cite{Transparency}, enterprise search systems facilitate productive and effective functioning of companies in obtaining information and help reduce search time by 15\% to 30\%. The statistics show that enterprise search systems play a critical role in corporate environment.




We propose an enterprise search pipeline on a real bilingual click dataset leveraging suitable state-of-the-art approaches and customise them according to our system needs. Then, we evaluate the customised ranking models with offline and online metrics to ensure high retrieval performance in a reasonable response time. Moreover, we mention the challenges of working on a real click-through dataset in enterprise domain. Then, we conclude the paper and give some future work.
\section{Proposed System}
\label{sec:system}
\begin{figure*}[!th]
  \centering  
  \includegraphics[width=\textwidth]{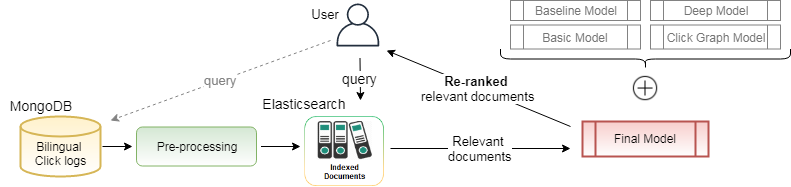}
  \caption{System Workflow}
  \label{fig:flowchart}
\end{figure*}

Our goal is to establish an end-to-end system for enterprise search on bilingual click-through dataset. To achieve this, we evaluate different ranking models on the dataset and customise these models when necessary to generate the combined model. Then, we construct the end-to-end pipeline by using the final model as denoted in Fig.~\ref{fig:flowchart}. In this section, we describe the proposed system with all the details. First in Sect.~\ref{sec:workflow}, we present the high-level workflow of the system and then discuss different ranking models in the following sections.

\subsection{System Workflow}
\label{sec:workflow}
The proposed system is composed of distinct models to leverage different type of information in the dataset. The dataset of bilingual click-through logs was initially stored in a MongoDB database. Then, we preprocessed the click-logs and indexed them in elasticsearch. The details about the preprocessing phase will be given in Sect.~\ref{sec:dataset}.

In our system, elasticsearch is used as a baseline search engine. As shown in the Fig.~\ref{fig:flowchart}, a user query issued to the system is firstly sent to elasticsearch and elasticsearch ranks and returns an initial set of relevant documents towards the given query. Subsequently, the initial set is re-ranked by our final model, i.e., which is established by combining all the features in the four distinct ranking models below, and the result set of documents are returned to the user. In this way, we  aim to decrease the user response time since we do not apply our final model on the whole corpus but rather on the initial set of the documents, i.e., generally the subset of the document corpus, returned by elasticsearch. 
In the overall system, our main target is to improve the retrieval performance of the baseline, elasticsearch and the assumption is that elasticsearch retrieves (almost) all the relevant documents in the corpus for the given query. Otherwise, our system cannot retrieve all the relevant documents towards the given query since its retrieval performance directly depends on the initial set of documents. Nonetheless, this is a natural accuracy-efficiency trade-off and a base search engine assumption is generally used in real search systems for efficiency purposes.

\subsection{Baseline Ranking Model}
\label{sec:baseline}
The first ranking model contains one feature which is BM25. We call this ranking model as "baseline" since elasticsearch uses the same feature, i.e., only BM25, for searching in the default case. We compare the following ranking models with this model to show an improvement over baseline.
\subsection{Basic Ranking Model}
\label{sec:basic}
The second model uses five tf*idf-variant (term frequency multiplied by inverse document frequency) features such as inverse document frequency (idf), document frequency (docfreq) etc. Please refer to the elasticsearch documentation for all the features that the elasticsearch use in tf*idf mode. These are the features computed by the elasticsearch if one chooses tf*idf metric for searching instead of the default BM25. Note that in the scope of the basic ranking model, one can include various text-matching features to the model as levenshtein distance, jaccard similarity etc. to improve the performance even more. We prefer to use only these five features in the model and there are two reasons for that. First, since we evaluate the online performance of our system as well in the Sect.~\ref{sec:experimental}, response time needs to be taken into consideration in addition to the retrieval performances of the ranking models, which constitute an offline evaluation. When our system sends a user query to the elasticsearch, elasticsearch computes all these five features in addition to BM25 even in the default search mode. Thus, our system obtains these already computed feature values from the elasticsearch efficiently in comparison to the following two ranking models. Second, we aim to make an offline evaluation on the retrieval performances of the different search modes of elasticsearch, i.e., BM25 and tf*idf, apart from the ranking models.

\begin{figure*}[!t]
  \centering  
  \includegraphics[width=0.8\textwidth]{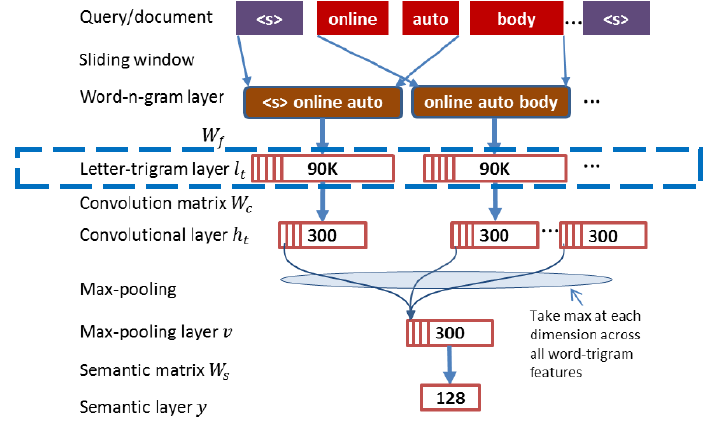}
  \caption{CLSM Architecture~\cite{shen2014latent}.}
  \label{fig:clsm}
\end{figure*}

\subsection{Deep Ranking Model}
\label{sec:deep}
The third model leverages a state-of-the-art deep neural network architecture specifically proposed for web search by Microsoft Research~\cite{shen2014latent}. The architecture is shown in Fig.~\ref{fig:clsm}. We use this model to find semantic relations between queries and documents. With the help of the model, the system can detect semantically relevant queries/documents which the previous two models containing text-matching features, may fail to discover. Thus, the deep ranking model strengthens the overall search system by providing the semantic search capability on top of keyword search.

Using this model, we generate the vector representations of queries and documents in semantic space from the click-through log which contains clicked and non-clicked query-document pairs. The assumption behind this model is that clicked query/document pairs' vector representations should be more similar than non-clicked ones and the network is trained based on this assumption. In the training phase, clicked pairs constitute positive samples and non-clicked ones, negative. Then, we compute a cosine similarity score for each query/document pair in the dataset and use this feature for the deep ranking model in retrieving the documents for a given query. Note that we customise the proposed deep model and remove the letter-trigram layer as denoted in Fig.~\ref{fig:clsm} to use it on the bilingual enterprise search dataset. The modification details will be validated in the experimental setup below.

\begin{figure*}[!t]
  \centering  
  \includegraphics[width=0.6\textwidth]{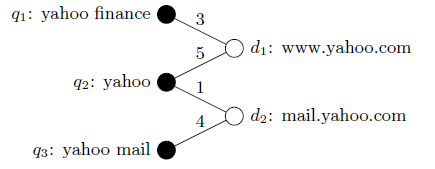}
  \caption{A sample of click-through bipartite graph~\cite{jiang2016learning}.}
  \label{fig:clickgraph}
\end{figure*}

\subsection{Click Graph Ranking Model}
\label{sec:clickgraph}
The final ranking model is inspired from the work of Yahoo Search~\cite{jiang2016learning}. The researchers propose a simple, yet effective idea of establishing a bipartite graph from only the clicked query/document pairs by denoting queries/documents as vertices and click information as edges where edge weights show click counts of the corresponding query-document pairs. A sample of click-through bipartite graph given in the original work is shown in Fig.~\ref{fig:clickgraph}.

After having constructed the graph from the clicked pairs, researchers represent queries (documents) in the query (document) space, i.e., only with words in the queries (documents), then propagate these vectors to the document (query) side by taking into account of the edge weights. The vector propagation algorithm is conducted until convergence. At the end of the algorithm, queries and documents are represented with the propagating words and the main aim is to denote the clicked queries and documents in the same semantic space through decreasing the lexical gap to alleviate noise and sparsity problems of click information. 
%

In addition to the idea of representing \emph{click-existing} queries/documents, the researchers also present a solution for \emph{click-absent} queries/documents, i.e., new queries/documents that are not in the click log, which is a critical problem for search systems. The main objective is to estimate the new queries/documents by associating them with the existing queries/documents that are already created by propagation on the click-graph. At this point, we propose to customise the two-step vector estimation algorithm of the \emph{click-absent} queries/documents. In the original paper, the absent queries/documents are connected with the existing ones in the graph by mainly looking at their common words, i.e., text-matching. After the connection, another vector propagation algorithm is executed on the graph by exploiting the connected queries/documents information in the graph and further apply an optimisation algorithm at the end to generate unit (unique ngram) vectors and their weights. This method probably fails to find the semantically related queries/documents in the graph through only leveraging text similarity without considering semantic matching, thus we believe that the vectors of their \emph{click-absent} counterparts would be estimated with rather low quality.

\begin{table}[!h]
  \begin{center}
    \caption{Features list of all ranking models.}
    \label{tab:features}
    \begin{tabular}{|l|c|c|} 
            \hline
      \textbf{Models} & \textbf{Features} & \textbf{Description}\\
            \hline
      \hline
      Baseline & F1 & Only BM25\\
            \hline
      Basic & F2-F6 & Five tf*idf variant features\\
            \hline
      Deep & F7 & Deep cosine similarity\\
        \hline
      Click Graph & F8-F9 & Click graph cosine \& jaccard similarity\\
        \hline
    \textbf{Final} & F1-F10 & All features\\
        \hline 
    \end{tabular}
  \end{center}
\end{table}

For creating relatively higher quality representations from the graph, we find semantically related queries/documents using word embeddings instead of looking at common words, then apply the vector propagation algorithm as in the original work using the click information to estimate the \emph{click-absent} representations. In this way, we also do not need an optimisation algorithms for the units. We use the Word2Vec word embeddings~\cite{mikolov2013efficient} to find semantically similar queries/documents in the graph. We then generate centroid vectors of \emph{click-absent} queries/documents and the existing ones in the graph using the embedding vectors suggested in~\cite{kuzi2016query}. Then, we find similar existing queries/documents with cosine similarity for each \emph{click-absent} query/document estimate their representations accordingly. After having generated or estimated the vector representations of all query/document pairs in the dataset, we compute two similarity scores between them as cosine \& jaccard similarity and use these two scores as features for the click graph ranking model. The features of each ranking model are displayed in Table~\ref{tab:features}.


\section{Experimental Setup}
\label{sec:experimental}
In this section, we provide a description of our experimental setup based on the ranking models as mentioned in Sect.~\ref{sec:system}. We firstly give details about the material and further mention the preprocessing steps applied on it before feeding the dataset to the system. Lastly, we make an offline/online evaluation of the system with several metrics.

\subsection{Dataset}
\label{sec:dataset}
The dataset is composed of clicked and non-clicked query and document pairs and crawled from an in-house search system of a company. The search system is similar to a ticket issue system where users can ask help about their technical problems and technical experts answer to these questions in the system. Then, other users experiencing similar problems can search and find relevant answers or solutions to their problem. 
The dataset contains mixed English/Chinese queries and documents together. Our dataset includes 466.000 query-document pairs where 146.000 of these are uniqe pairs. For each query-document pair, we have the following information: user id, query, document title, document url, clicked or non-clicked. 

We note that the system is less dynamic in comparison to a web search engine or publicly available search system since it is an in-house search platform, thus its number of users is limited. Moreover, the system works in a technical domain, thereby containing a rather limited vocabulary of technical terms which in a sense makes our problem easier. Yet, queries and documents include mixed English and Chinese words at the same time which has its own challenges. Chinese language is very different than English such that new words are not constituted with letters in the alphabet but using different symbols; one cannot tokenise Chinese content like English. Thus we had to remove the letter-trigram layer from the original CLSM architecture as denoted in Fig.~\cite{shen2014latent}. This layer may not be necessary to our system as in the case of Microsoft web commercial engine since we do not have a big vocabulary size. These differences led us to modify the preprocessing phase and also customise the ranking models accordingly.

\subsection{Preprocessing}
\label{sec:preprocessing}
In the preprocessing phase, we firstly found the unique query-document pairs by accumulating the click counts without looking at user information since in the scope of this work, we did not aim to make a user analysis. This is because, the dataset did not have sufficient search information of a specific user or a user segment for the analysis. Second, we removed punctuations and stop words in English and Chinese. Then, we made the queries and documents lowercased. After that, we tokenised English with space/tab delimiters and segmented Chinese content using Stanford NLP Parser~\cite {manning2014stanford}. We further trained an offline Word2Vec CBOW model~\cite{mikolov2013efficient} on the corpus to find semantically similar queries and documents to be used for click-absent query/document vector estimation.

\subsection{Evaluation Results}
\label{sec:Evaluation}
We firstly made an offline evaluation and evaluated the retrieval performance of different (customised if necessary) ranking models mentioned above using eight well-known ranking algorithms from RankLib~\cite{danglemur} which is a library of learning to rank algorithms and easy to use. We have two offline evaluation metrics as Normalised Discounted Cumulative Gain (NDCG) at ranks 5 and 10 on 5-fold cross-validation.

We firstly experimented with four different network configurations of the deep rank model without the letter-trigram layer as mentioned in Sect.~\ref{sec:deep}. To find the best parameter configuration, we used NDCG@10 metric and then integrated the optimum configuration to the final pipeline. The results are shown in Table~\ref{tab:deep}. The best score of each configuration is highlighted as bold and the results show that the best configuration is the \emph{Deep Model1}. We used that model for the deep ranking model part of the final pipeline. Then, we compared the retrieval performances of five ranking models including the \emph{Final model} with NDCG@5 and NDCG@10 metrics and the results are shown in 
Table~\ref{tab:ndcg5} and Table~\ref{tab:ndcg10} respectively.

\begin{table}[!h]
  \begin{center}
    \caption{NDCG@10 evaluation scores of different network configurations of deep ranking models on 5-fold cv.}
    \label{tab:deep}
    \resizebox{\columnwidth}{!}{%
    \begin{tabular}{l|c|c|c|c|c|c|c|c|} 
      & \textbf{MART} & \textbf{RankNet} & \textbf{RankBoost} & \textbf{AdaRank} & \textbf{Coor.Ascent} & \textbf{$\lambda$-Rank} & \textbf{$\lambda$-MART} & \textbf{ListNet} \\
            \hline
      \hline
      \textbf{Deep Model1} & 0.9128 & 0.9116 & \textbf{0.9230} & 0.9116  & 0.9116 & 0.8957 & 0.9122 & 0.9116 \\
            \hline
      Deep Model2 & 0.8994 & 0.8987 & \textbf{0.9109} & 0.8987  & 0.8987 & 0.8919 &0.8984 & 0.8987\\
            \hline
      Deep Model3 & 0.8995 & 0.8992 & \textbf{0.9091} & 0.8992  &0.8992 & 0.8938 & 0.8977 & 0.8992\\
        \hline
      Deep Model4 & 0.8986 & 0.8989 & \textbf{0.9098} & 0.8989   & 0.8989 & 0.8817 & 0.8980 & 0.8989\\
    \hline
    \end{tabular}
    }
  \end{center}
\end{table}

\begin{table}[!h]
  \begin{center}
    \caption{NDCG@5 evaluation scores on 5-fold cv and online response times of the ranking algorithms.}
    \label{tab:ndcg5}
    \resizebox{\columnwidth}{!}{%
    \begin{tabular}{l|c|c|c|c|c|c|c|c|} 
      & \textbf{MART} & \textbf{RankNet} & \textbf{RankBoost} & \textbf{AdaRank} & \textbf{Coor.Ascent} & \textbf{$\lambda$-Rank} & \textbf{$\lambda$-MART} & \textbf{ListNet} \\
            \hline
      \hline
      Baseline & 0.9048 & 0.9012 & \textbf{0.9239} & 0.9068  & 0.9068 & 0.8992 & 0.9082 & 0.9068 \\
            \hline
      Basic & 0.8975 & 0.8943 & \textbf{0.9079} & 0.9085  & 0.8952 & 0.8959 &0.9002& 0.8939\\
            \hline
      Deep & 0.9128 & 0.9116 & \textbf{0.9230} & 0.9116  &0.9116 & 0.8957& 0.9122& 0.9116\\
        \hline
      Click Gr. & 0.9301 & 0.9178 & 0.9307 & 0.9180   & 0.9192& 0.9128& 0.9315& 0.9179\\

        \hline \\
        \hline 
        \hline
        Res. Time & 0.11076 & 0.01237 & 0.01093 & 0.01073  &0.01126 & 0.10958& 0.02735& 0.01239\\
                \hline
    \end{tabular}
    }
  \end{center}
\end{table}

\begin{table}[!h]
  \begin{center}
    \caption{NDCG@10 evaluation scores on 5-fold cv and online response times of the ranking algorithms.}
    \label{tab:ndcg10}
    \resizebox{\columnwidth}{!}{%
    \begin{tabular}{l|c|c|c|c|c|c|c|c|} %
      & \textbf{MART} & \textbf{RankNet} & \textbf{RankBoost} & \textbf{AdaRank} & \textbf{Coor.Ascent} & \textbf{$\lambda$-Rank} & \textbf{$\lambda$-MART} & \textbf{ListNet} \\
            \hline
      \hline
      Baseline & 0.9048 & 0.9012 & 0.9239 & 0.9068  & 0.9068 & 0.8992 & 0.9082 & 0.9068 \\
            \hline
      Basic & 0.8975 & 0.8943 & 0.9079 & 0.9085  & 0.8952 & 0.8959 &0.9002& 0.8939\\
            \hline
      Deep & 0.9128 & 0.9116 & 0.9230 & 0.9116  &0.9116 & 0.8957& 0.9122& 0.9116\\
        \hline
      Click Gr. & 0.9301 & 0.9178 & 0.9307 & 0.9180   & 0.9192& 0.9128& 0.9315& 0.9179\\
        \hline
    \textbf{Final} & 0.9309 & 0.9027 & 0.9295 & 0.9188  &0.9204 & 0.8954& 0.9315& 0.9119\\
        \hline \\
        \hline
        \hline
        Res. Time & 0.11076 & 0.01237 & 0.01093 & 0.01073  &0.01126 & 0.10958& 0.02735& 0.01239\\
                \hline
    \end{tabular}
    }
  \end{center}
\end{table}

\newpage



\section{Conclusion \& Future Work}
In this paper, we reported our results on the real bilingual dataset in technical domain. To establish an end-to-end system, we benefited from the approaches proposed in literature. Then, we combined those approaches to improve the search results in an in-house and domain-specific search platform which returns only technical documents. We believe that this report is valuable since it describes our experience of applying proposed approaches in production environment.

\bibliographystyle{splncs03}
\bibliography{references}

\begin{thebibliography}{10}
\providecommand{\url}[1]{\texttt{#1}}
\providecommand{\urlprefix}{URL }

\bibitem{EnterpriseSearch}
Enterprise search.
  \url{http://david-hawking.net/pubs/ModernIR2\_Hawking\_chapter.pdf},
  accessed: 2019-11-06

\bibitem{Transparency}
Enterprise search market - global industry analysis, size, share, growth,
  trends, and forecast 2013 - 2019.
  \url{https://pitchengine.com/pitches/76c4358e-19ab-4113-b524-f00c3b98de29},
  accessed: 2018-12-10

\bibitem{InternetLiveStats}
Internetlivestats. \url{http://www.internetlivestats.com/}, accessed:
  2018-10-06

\bibitem{danglemur}
Dang, V.: The lemur project-wiki-ranklib. Lemur Project,[Online]. Available:
  http://sourceforge. net/p/lemur/wiki/RankLib

\bibitem{jiang2016learning}
Jiang, S., Hu, Y., Kang, C., Daly~Jr, T., Yin, D., Chang, Y., Zhai, C.:
  Learning query and document relevance from a web-scale click graph. In:
  Proceedings of the 39th International ACM SIGIR conference on Research and
  Development in Information Retrieval. pp. 185--194. ACM (2016)

\bibitem{kuzi2016query}
Kuzi, S., Shtok, A., Kurland, O.: Query expansion using word embeddings. In:
  Proceedings of the 25th ACM international on conference on information and
  knowledge management. pp. 1929--1932. ACM (2016)

\bibitem{manning2014stanford}
Manning, C., Surdeanu, M., Bauer, J., Finkel, J., Bethard, S., McClosky, D.:
  The stanford corenlp natural language processing toolkit. In: Proceedings of
  52nd annual meeting of the association for computational linguistics: system
  demonstrations. pp. 55--60 (2014)

\bibitem{mikolov2013efficient}
Mikolov, T., Chen, K., Corrado, G., Dean, J.: Efficient estimation of word
  representations in vector space. arXiv preprint arXiv:1301.3781  (2013)

\bibitem{SpendTime}
Noi, D.D.: Do workers still waste time searching for information?
  \url{https://blog.xenit.eu/blog/do-workers-still-waste-time-searching-for-information},
  accessed: 2019-11-06

\bibitem{shen2014latent}
Shen, Y., He, X., Gao, J., Deng, L., Mesnil, G.: A latent semantic model with
  convolutional-pooling structure for information retrieval. In: Proceedings of
  the 23rd ACM international conference on conference on information and
  knowledge management. pp. 101--110. ACM (2014)

\end{thebibliography}
\end{document}